\documentclass[preprint,showpacs,preprintnumbers,amsmath,amssymb]{revtex4}
\usepackage{graphicx}
\usepackage{dcolumn}
\usepackage{bm}

\begin{document}

\title{Fermionic zero modes in self-dual vortex background on a
torus\footnote{ The project supported by National Natural Science
Foundation of China under Grant No. 10275030 and 10475034 and the
Fundamental Research Fund for Physics and Mathematic of Lanzhou
University (No. Lzu07002).} }

\author{Yu-Xiao Liu$^1$\footnote{Corresponding author. Email: liuyx@lzu.edu.cn},
 Yong-Qiang Wang$^2$ and Yi-Shi Duan$^1$}

\affiliation{$^1$Institute of Theoretical Physics, Lanzhou
University, Lanzhou 730000, China\\
$^2$Zhejiang Institute of Modern Physics, Department of Physics,
Zhejiang University, Hangzhou 310027, China }

\begin{abstract}
We study fermionic zero modes in the self-dual vortex background
on an extra two-dimensional Riemann surface in 5+1 dimensions.
Using the generalized Abelian Higgs model, we obtain the inner
topological structure of the self-dual vortex and establish the
exact self-duality equation with topological term. Then we analyze
the Dirac operator on an extra torus and the effective Lagrangian
of four-dimensional fermions with the self-dual vortex background.
Solving the Dirac equation, the fermionic zero modes on a torus
with the self-dual vortex background in two simple cases are
obtained.
\end{abstract}

\pacs{11.10.Kk, 04.50.+h\\
Key words: Fermionic zero modes, large Extra Dimensions, self-dual
vortex}

\maketitle

\section{Introduction}

In four-dimensional space-time, the interactions of fermions in a
Nielsen-Olesen vortex background have been widely analyzed in the
literature, mainly in connection with bound states at threshold
\cite{NohlPRD1975}, zero modes \cite{JackiwPRD1984} and scattering
solutions \cite{VegaPRD19787}. In Ref. \cite{StojkovicPRD2001},
Stojkovic $et\;al$ discussed some interesting situations arising
in cases when fermions can have a non-trivial mass matrix. Such
situations can also arise in higher dimensional models and can
change the conclusions of fermion localization on a vortex
(existence and non-existence of fermionic zero modes). Recently,
Frere, Libanov and Troitsky have shown that a single family of
fermions in six dimensions with vector-like couplings to the
Standard Model (SM) bosons gives rise to three generations of
chiral SM fermions in four dimensions \cite{LibanovNPB2001599}. In
5+1 dimensions, Frere $et\;al$ also studied the fermionic zero
modes in the background of a vortex-like solution on an extra
two-dimensional sphere and relate them to the replication of
fermion families in the SM \cite{FrereJHEP20030306}.

The topological vortex (especially Abrikosov-Nielsen-Olesen
vortex) coupled to fermions may lead to chiral fermionic zero
modes \cite{JackiwRossiNPB1981}. Usually the number of the zero
modes coincides with the topological number, that is, with the
magnetic flux of the vortex. In Large Extra Dimensions (LED)
models, the chiral fermions of the SM are described by the zero
modes of multi-dimensional fermions localized in the
(four-dimensional) core of a topological defect
\cite{RemarkableApplication}. One spinor field of theory in 5+1
dimensions corresponds to three chiral fermions of effective
theory in 3+1 dimensions. Due to this fact, number of free
parameters of the model can be significantly reduced.

In Refs. \cite{hep-th/0508104, ourMPLA2005}, we present a unified
description of the topological and non-topological self-dual
vortices on a two-dimensional non-compact extra space. Based on
this vortex background, we study fermionic zero modes on the extra
space in 5+1 dimensions \cite{ourMPLA2005}. Through two simple
cases, it is shown that the vortex background contributes a phase
shift to the fermionic zero modes. The phase is actually
originated from the Aharonov-Bohm (AB) effect and can be divided
into two parts, one is related with the topological number of the
extra space, the other depends on the non-topological vortex
solution. While the model of 5+1 dimensions with two extra compact
dimensions can provide an interesting insight into the problem of
hierarchy problem of chiral fermionic mass pattern. In this paper,
we shall study fermionic zero modes coupled with a self-dual
vortex background on two-dimensional compact extra spaces $T^2$ in
5+1 dimensions.

The paper is organized as follows: In section \ref{SectionVortex},
through the self-duality equation on a two-dimensional curved
space, we give the topological structure of self-dual vortex. In
section \ref{OnT2}, we analyze the Dirac operator and the
effective lagrangian of the fermions in the self-dual vortex
background on a torus in 5+1 dimensions, and two simple cases are
discussed to show the role of vortex background in the fermionic
zero modes. In the last section, a brief conclusion is presented.

\section{Self-dual vortex on a two-dimensional curved Riemann surface}
\label{SectionVortex}

We consider a (5+1)-dimensional space-time $M^{4}\times{K^2}$ with
$M^{4}$ represents our four-dimensional space-time and $K^2$
represents a two-dimensional extra Riemann surface. The metric
$G_{MN}$ of the manifold $M^{4}\times{K^2}$ is determined by
\begin{eqnarray}
ds^2=G_{MN}dx^M dx^N =g_{\mu \nu}dx^\mu{d}x^\nu-
\gamma_{ij}dy^{i}dy^{j}, \label{metric}
\end{eqnarray}
where $g_{\mu \nu}=g_{\mu \nu}(x)$ is the four-dimensional metric
of the manifold $M^4$, $\gamma_{ij}=\gamma_{ij}(y)$ is the
two-dimensional metric of the extra space $K^2$.

To generate the vortex solution, we introduce the Abelian Higgs
Lagrangian
\begin{equation}
\mathcal{L}_{AH}=\sqrt{-G} \left
(-\frac{1}{4}F_{MN}F^{MN}+(D^{M}\phi)^{\dag}(D_{M}\phi)-\frac{\lambda}{2}(\|\phi\|^{2}-v^{2})^{2}
\right ),\label{lagrangianAH}
\end{equation}
where $G=\det(G_{MN})$,
$F_{MN}=\partial_{M}A_{N}-\partial_{N}A_{M}$, $\phi=\phi(y^k)$ is
a complex scalar field on $K^2$,
$\|\phi\|=(\phi\phi^{\ast})^{\frac{1}{2}}$, $A_{M}$ is a U(1)
gauge field, $D_{M}=\partial_{M}-ieA_{M}$ is gauge-covariant but
not covariant under general coordinate transformation. The
Abrikosov-Nielsen-Olesen vortex solution on the $M^{4}\times{K^2}$
could be generated from the Higgs field. In the generalized
Abelian Higgs model, if the system admits a Bogomol'nyi limit
\cite{BogoSJNP1976}, one can arrive at the first-order Bogomol'nyi
self-duality equations in a curved space-time \cite{KimJMP2002}:
\begin{eqnarray}
&&B=\mp e(\|\phi\|^{2}-v^2),\label{bogo1}\\
&&D_{i}\phi\mp{i}\sqrt{\gamma}\epsilon_{ij}\gamma^{jk}D_{k}\phi=0.
\label{bogo2}
\end{eqnarray}
The complex Higgs field $\phi$ can be regarded as the complex
representation of a two-dimensional vector field
$\vec{\phi}=(\phi^{1}, \phi^{2})$ over the base space, it is
actually a section of a complex line bundle on the base manifold.
Substituting $\phi=\phi^{1}+i\phi^{2}$ and
$D_{i}=\partial_{i}-ieA_{i}$ into Eq. (\ref{bogo2}) and splitting
the real part form the imaginary part, we obtain two equations
\begin{eqnarray}
&&\partial_{i}\phi^{1}=-eA_{i}\phi^{2}
\mp\sqrt{\gamma}\epsilon_{ij}\gamma^{jk}(+\partial_{k}\phi^{2}
-eA_{k}\phi^{1}),\label{split1}\\
&&\partial_{i}\phi^{2}=+eA_{i}\phi^{1}
\mp\sqrt{\gamma}\epsilon_{ij}\gamma^{jk}(-\partial_{k}\phi^{1}
-eA_{k}\phi^{2}).\label{split2}
\end{eqnarray}
From Eqs. (\ref{split1}) and (\ref{split2}), by calculating
$\partial_{i}\phi^{*}\phi-\partial_{i}\phi\phi^{*}$, we can obtain
the expression of the gauge potential
\begin{eqnarray}
eA_{i}=-\frac{1}{2i\|\phi\|^{2}}(\partial_{i}\phi^{*}\phi-\partial_{i}\phi\phi^{*})
\mp \sqrt{\gamma}\epsilon_{ij}\gamma^{jk}\partial_{k}\ln\|\phi\|.
\label{eAi1}
\end{eqnarray}
If we define the unit vector
\begin{equation}\label{na}
n^{a}=\frac{\phi^{a}}{\|\phi\|}, \;\;\;\;(a,b=1,2)
\end{equation}
and note the identity
\begin{equation}\label{identity}
\epsilon_{ab}n^{a}\partial_{i}n^{b}=\frac{1}{2i\|\phi\|^{2}}
(\partial_{i}\phi^{*}\phi-\partial_{i}\phi\phi^{*}),
\end{equation}
Eq. (\ref{eAi1}) further simplifies to:
\begin{equation}
eA_{i}=-\epsilon_{ab}n^{a}\partial_{i}n^{b}
\mp\sqrt{\gamma}\epsilon_{ij}\gamma^{jk}\partial_{k}\ln\|\phi\|.\label{eAi2}
\end{equation}
In curved space, the magnetic field is defined by
$B=-\frac{1}{\sqrt{\gamma}}\epsilon^{ij}\partial_{i}{A}_{j}$,
according to Eq. (\ref{eAi2}), we have
\begin{equation}\label{B=delta+ln}
e\sqrt{\gamma}B=\epsilon^{i
j}\epsilon_{ab}\partial_{i}n^{a}\partial_{j}n^{b}
\pm\epsilon^{ij}\epsilon_{jk}\partial_{i}(\sqrt{\gamma}\gamma^{kl}\partial_{l}\ln\|\phi\|).
\end{equation}
So the first self-duality equation (\ref{bogo1}) can be
generalized to
\begin{equation}
\mp e^2 \sqrt{\gamma} (\|\phi\|^{2}-v^2) =
\epsilon^{ij}\epsilon_{ab}\partial_{i}n^{a}\partial_{j}n^{b}
\pm\epsilon^{ij}\epsilon_{jk}\partial_{i}(\sqrt{\gamma}\gamma^{kl}\partial_{l}\ln\|\phi\|).\label{nonlinear01}
\end{equation}
According to Duan's $\phi$-mapping topological current theory
\cite{DuanKT1976}, it is easy to see that the first term on the
RHS of Eq. (\ref{nonlinear01}) bears a topological origin, and the
topological term just describes the non-trivial distribution of
$\vec{n}$ at large distances in space \cite{'tHooft}. Noticing
$\partial_{i}n^{a}={\partial_{i}\phi^{a}}/{\parallel\phi\parallel}
+\phi^{a}\partial_{i}({1}/{\parallel\phi\parallel})$ and the Green
function relation in $\phi$-space:
$\partial_{a}\partial_{a}ln(\|\phi\|)=2\pi\delta^{2}(\vec{\phi}\;),\;
(\partial_{a}={\frac{\partial}{\partial\phi^{a}}})$, it can be
proved that \cite{DuanNPB1998514}
\begin{equation}
\epsilon^{i j}\epsilon_{ab}\partial _{i}n^{a}\partial_{j}n^{b}
=2\pi\delta^{2}(\vec{\phi}\;)J(\frac{\phi}{y})
=2\pi\sum_{k=1}^NW_k\delta (\vec{y}-\vec{y_{k}}), \label{BT=del}
\end{equation}
where $J(\phi/y)$ is the Jacobian and $W_k=\beta _k\eta _k$ is the
winding number around the $k$-th vortex, the positive integer
$\beta_k$ is the Hopf index and $\eta _k=\pm 1$ is the Brouwer
degree, $\vec{y_{k}}$ are the coordinates of the $k$-th vortex. So
the first Bogomol'nyi self-duality equation (\ref{bogo1}) should
be
\begin{eqnarray}
\mp e^2 \sqrt{\gamma} (\|\phi\|^{2}-v^2)
=2\pi\sum_{k=1}^NW_k\delta (\vec{y}-\vec{y_{k}}) ~\mp~
\epsilon^{ij}\epsilon_{jk} \partial_{i} (\sqrt{\gamma} \gamma^{kl}
\partial_{l} \ln\|\phi\|).\label{generaleq}
\end{eqnarray}
Obviously the first term on the RHS of Eq. (\ref{generaleq})
describes the topological self-dual vortex.

Now let us discuss the case of flat space for the self-duality
equation (\ref{generaleq}). In this special case,
$\gamma_{ij}=\delta_{ij}$ and Eq. (\ref{generaleq}) reads as
\begin{equation}
\mp e^2(\|\phi\|^{2}-v^2) %
=2\pi\sum_{k=1}^NW_k\delta (\vec{y}-\vec{y_{k}}) %
~\mp~\partial_{i}\partial_{i}\ln\|\phi\|.\label{flateq}
\end{equation}
While the corresponding conventional self-duality equation is
\cite{Dunne1998}
\begin{equation}
e^2(\|\phi\|^{2}-v^2)
=\partial_{i}\partial_{i}\ln\|\phi\|.\label{conventionalEq}
\end{equation}
Comparing our equation (\ref{flateq}) with Eq.
(\ref{conventionalEq}), one can see that the topological term $ 2
\pi {\sum_{k=1}^N} {W_k} \delta (\vec{y} - \vec{y_{k}})$, which
describes the topological self-dual vortex, is missed in the
conventional equation. Obviously, only when the field $\phi\neq0$,
the topological term vanishes and the conventional equation is
correct. So, the exact self-duality equation should be Eq.
(\ref{flateq}) for flat space and Eq. (\ref{generaleq}) for curved
one. As for conventional self-dual nonlinear equation
(\ref{conventionalEq}), a great deal of work has been done by many
physicists on it, and a vortex-like solution was given by Jaffe
\cite{jaffe}. But no exact solutions are known.

For the case of a flat torus, $\gamma_{ij}dy^{i}dy^{j}=R_1^2
d\theta^2 + R_2^2 d\varphi^2$, the corresponding self-dual vortex
equation is
\begin{eqnarray}
\mp e^2 R_1 R_2 (\|\phi\|^{2}-v^2) =2\pi\sum_{k=1}^NW_k\delta
(\vec{y}-\vec{y_{k}}) ~\mp~ \left(
\frac{R_2}{R_1}\partial_{\theta}^2 \ln\|\phi\| +
\frac{R_1}{R_2}\partial_{\varphi}^2
\ln\|\phi\|\right).\label{T2VortexEq}
\end{eqnarray}
For the case of a sphere, $\gamma_{ij}dy^{i}dy^{j}=R^2(d\theta^2 +
\sin^2 \theta d\varphi^2)$, and we have
\begin{equation}
\mp e^2 R^2 \sin \theta (\|\phi\|^{2}-v^2)
=2\pi\sum_{k=1}^NW_k\delta (\vec{y}-\vec{y_{k}}) %
~~\mp~ \left( \partial_{\theta}(\sin\theta \partial_{\theta}
\ln\|\phi\|) + \frac{1}{\sin\theta}\partial_{\varphi}^2
\ln\|\phi\|\right).\label{S2VortexEq}
\end{equation}
We shall study fermionic zero modes coupled with the vortex
background on a torus in the following section.

\section{Fermionic zero modes in vortex background on a torus}
\label{OnT2}

The lagrangian of the fermions in the vortex background
(\ref{eAi2}) on a torus is
\begin{equation}
\mathcal{L}=\sqrt{-G} \bar{\Psi}\{ i \Gamma^A E^M_A (\partial_M -
\Omega_M -ieA_M)-g\phi\}\Psi, \label{LpsiT2}
\end{equation}
where $E^{M}_{A}$ is the {\sl sechsbein} with
\begin{equation} \label{sechsbein}
E^{A}_{M} = \left(e^{a}_{\mu}
\delta^{A}_{a},R_1\delta^{A}_{4},R_2\delta^{A}_{5}\right).
\end{equation}
The 
components of $\Omega_M$ are
\begin{equation}
\Omega_{\mu}=\omega_{\mu},~~~\Omega_{4}=0,~~~\Omega_{5}=0,\label{connectionT2}
\end{equation}
where $\omega_{\mu}=\frac{1}{2} \omega_{\mu}^{ab}I_{ab}$ is the
spin connection derived from the metric $g_{\mu\nu}(x)=e_{\mu}^{a}
e_{\nu}^{b} \eta_{ab}$, lower case Latin indices $a,b=0,\cdots,3$
correspond to the flat tangent four-dimensional Minkowski space.

Using Eq. (\ref{connectionT2}), the lagrangian (\ref{LpsiT2}) of
the fermions then becomes
\begin{eqnarray}
\mathcal{L}&=&\sqrt{-G} \bar{\Psi} \left\{ i \Gamma^a e^{\mu}_{a}
(\partial_{\mu} - \omega_{\mu} -ie A_{\mu})+  i
\frac{\Gamma^4}{R_1}(\partial_{\theta} - ie A_{\theta})\right. \nonumber \\
&& \left. + i\frac{\Gamma^5}{R_2} (\partial_{\varphi} -
ieA_{\varphi} ) - g \phi \right\} \Psi.\label{LpsiT22}
\end{eqnarray}
We apply now the standard decomposition procedure. Since the
vortex background does not depend on $x^{\mu}$, one can separate
variables related to $M^{4}$ and $T^{2}$. First, let us introduce
the transverse Dirac operator  $D_T$ on a torus in the background
(\ref{eAi2}):
\begin{eqnarray}
D_T = i\frac{\bar{\Gamma}\Gamma^4 }{R_1}(\partial_{\theta} - ie
A_{\theta})
 + i\frac{\bar{\Gamma}\Gamma^5}{R_2} \left ( \partial_{\varphi}
-ie A_{\varphi} \right )-\bar{\Gamma}g\phi,
\end{eqnarray}
where
\begin{equation}
\bar{\Gamma}=\Gamma^0 \Gamma ^1 \Gamma ^2\Gamma ^3=\left(%
\begin{array}{cc}
  i\gamma^5 & 0 \\
  0 & -i\gamma^5 \\
\end{array}%
\right).
\end{equation}
Then expand any spinor in a set of eigenvectors
$\Theta_m(\theta,\varphi)$ of this operator $D_T$
\begin{equation}
D_T\Theta_m(\theta,\varphi)=\lambda_m\Theta_m(\theta,\varphi).
\end{equation}
For the zero modes of $D_T$, we have
\begin{equation}
D_T\Theta(\theta,\varphi)=0. \label{DiracEqofDT}
\end{equation}
This is just the Dirac equation on a torus with vortex
backgrounds. For fermionic zero modes, we can write
\begin{equation}
\Psi(x,\theta,\varphi)=\psi(x)\Theta(\theta,\varphi),
\end{equation}
where $\Theta$ satisfies Eq. (\ref{DiracEqofDT}). The effective
Lagrangian for $\psi$ then becomes
\begin{eqnarray}
{\cal L}_{eff} = %
\int d\theta d\varphi \;{\cal L} = {\cal L}_{\psi}\int d\theta
d\varphi R_1 R_2 \Theta^\dag \Theta,
\end{eqnarray}
where
\begin{eqnarray}
{\cal L}_{\psi} = %
\sqrt{-\det(g_{\mu\nu})} \; i \bar{\psi} \Gamma^a e^{\mu}_a
(\partial_{\mu} - \omega_{\mu} -ieA_{\mu}) \psi.
\end{eqnarray}
Thus, to have the finite kinetic energy for $\psi$, the above
integral must be finite. This can be achieved if the function
$\Theta(\theta,\varphi)$ does not diverge on the whole torus.

In what follows, to illustrate how the vortex background affects
the fermionic zero modes, we first discuss the simple case that
the Higgs field $\phi$ is only relative to $\varphi$, and then
solve the general Dirac equation for the vacuum Higgs field
solution $\|\phi\|=v$.

\subsection{Case I: $\phi$ is only relative to $\varphi$}

\indent\indent Now we discuss a simple situation that $\phi$ only
depends on the parameter $\varphi$, i.e., $\phi=\phi(\varphi)$. In
this case, Eq. (\ref{eAi2}) reduces to:
\begin{eqnarray}
A_{\theta}&=&\mp\frac{1}{e}\frac{R_{1}}{R_2}
\partial_{\varphi}\ln\|\phi\|,\label{A1CaseI}\\
A_{\varphi}&=&-\frac{1}{e}\epsilon_{ab}n^{a}
\partial_{\varphi}n^{b},\label{A2CaseI}
\end{eqnarray}
and Dirac equation $D_T\Theta=0$ becomes:
\begin{eqnarray} \label{DTCaseI}
i\bar{\Gamma}\left\{\frac{\Gamma^4}{R_1}\partial_{\theta}
+\frac{\Gamma^5}{R_2}\left(\partial_{\varphi}
-ie\frac{R_{2}}{R_1}A_{\theta}\Gamma^4\Gamma^{5}-ie A_{\varphi}
\right )+ig\phi\right\}\Theta(\theta,\varphi)=0.
\end{eqnarray}
Here $\Theta$ can be written as the following form:
\begin{equation}
\Theta(\theta,\varphi)=\exp \left[ i(n+\frac{1}{2})\theta \right]
h(\varphi),
\end{equation}
where $n$ is an integer, and $h(\varphi)$ satisfies
\begin{eqnarray} \label{hICaseI}
\left\{\partial_{\varphi}
+i\frac{R_{2}}{R_1}\left(n+\frac{1}{2}-eA_{\theta}\right)\Gamma^4\Gamma^{5}-ieA_{\varphi}
-ig{R_2}\phi\Gamma^5\right\}h(\varphi)=0.
\end{eqnarray}
Substituting Eq. (\ref{A1CaseI}) into Eq. (\ref{hICaseI}), we get
\begin{eqnarray}\label{hIICaseI}
\left\{ \partial_{\varphi}
-\left[\frac{R_{2}}{R_1}\left(n+\frac{1}{2}\right) \pm
\partial_{\varphi}\ln\|\phi\| \right]
\left(%
\begin{array}{cc}
  \gamma^5 & 0 \\
  0 & \gamma^5 \\
\end{array}%
\right) \right. \nonumber\\
\left. -ieA_{\varphi} -ig{R_2}\phi
\left(%
\begin{array}{cc}
  0 & \gamma^0 \\
  -\gamma^0 & 0 \\
\end{array}%
\right)
\right\}h(\varphi)=0~~~~~~~
\end{eqnarray}
with
\begin{eqnarray} \label{h'h}
h(\varphi)=
\left(%
\begin{array}{c}
  h_1 (\varphi) \\
  h_2 (\varphi) \\
\end{array}
\right).
\end{eqnarray}
It is very difficult to solve explicitly the above Dirac equation.
Let us consider the case of small extra dimensions, i.e., $R_1,R_2
\ll 1$. Under this assumption, we can ignore the last item on the
LHS of Eq. (\ref{hIICaseI}). And imposing the chirality condition
$\gamma^5 h_i(\varphi) = +h_i (\varphi),\;(i=1,2)$, Eq.
(\ref{hIICaseI}) can be simplified as
\begin{eqnarray}\label{hiCaseI}
\left\{\partial_{\varphi}
-\left[\frac{R_{2}}{R_1}\left(n+\frac{1}{2}\right) \pm
\partial_{\varphi}\ln\|\phi\| +ieA_{\varphi} \right]\right\}h_i
(\varphi)=0.
\end{eqnarray}
So the fermionic zero mode is
\begin{eqnarray}
\Theta(\theta,\varphi)&=& \|\phi\|^{\pm 1}
\exp\left[i\left(n+\frac{1}{2}\right)\theta+\frac{R_{2}}{R_1}\left(n+\frac{1}{2}\right)\varphi
\right.\nonumber\\
&&+\left. ie \int d\varphi A_{\varphi} \right]
\left(%
\begin{array}{c}
  1 \\
  1 \\
\end{array}%
\right) \otimes
\left(%
\begin{array}{r}
  1 \\
  0 \\
  1 \\
  0 \\
\end{array}%
\right).\label{modes1}
\end{eqnarray}
Now we see, the gauge potential $A_{\varphi}$ appears as phase
factor to the fermionic zero modes. So the phase factor is decided
by the gauge potential $A_{\varphi}$, but has no relation with
$A_{\theta}$. And $A_{\varphi}$ has no contribution to effective
Lagrangian, but $A_{\theta}$ does. Furthermore, for the
anti-vortex, the corresponding zero mode is singular at the zero
of $\phi$.
It suggests that the symmetric phase just corresponds to
singularity of fermionic zero mode, the topological self-dual
vortex just arises from the symmetric phase of Higgs field. So
this singularity must bear a topological origin, it is determined
by the topology of the physical system.

As all known, quantum topological and geometrical phases are
ubiquitous in modern physics---in cosmology, particle physics,
modern string theory and condensed matter. In fact, according to
Eq. (\ref{modes1}), we see this phase shift is actually the
quantum mechanical AB phase. This discussion can be generalized to
the AB phase of non-Abelian gauge theories, such as the Wilson and
't Hooft loops. Since the AB phase is fundamental to theories of
anyons and to gauge fields, it is an important tool for studying
the issues of confinement and spontaneous symmetry breaking.

\subsection{Case II: the vacuum solution}

\indent\indent In this subsection, we choose polar coordinates
$(r,\chi)$ which origin is at the center of vortex, and discuss
the fermions around the vortex for the case of the vacuum
solution. The relation of $(r,\chi)$ and $(\theta,\varphi)$ is
\begin{equation}
R_1 \theta =y^1 =r \cos\chi, \;\;\; R_2 \varphi =y^2 =r \sin\chi.
\end{equation}
In the case, the gauge is decided by
\begin{eqnarray}
ds^2&=g_{\mu \nu}dx^\mu{d}x^\nu- dr^2 - r^2 d\chi^2.
\end{eqnarray}
{\sl Sechsbein} $E^{A}_{M}$  and the non-zero components of
$\Omega_M$ are
\begin{eqnarray}
&&E^{A}_{M} = \left(e^{a}_{\mu} \delta^{A}_{a},\delta^{A}_{4},r\delta^{A}_{5}\right),\\
\label{sechsbeinT2II}
&&\Omega_{\mu}=\omega_{\mu}, \; \Omega_{5}=\frac{1}{2} \Gamma^4
\Gamma^5,\label{connectionT2II}
\end{eqnarray}
For the vacuum solution, $\|\phi\|^2=v^2$, i.e. $\phi=ve^{i\chi},
\; n^{r}=\cos\chi,\; n^{\chi}=\sin\chi$, we have $eA_{r}=0,\;
eA_{\chi}=-1$. Dirac equation $D_T \Theta=0$ is
\begin{eqnarray} \label{DTCaseII}
i\bar{\Gamma}\left\{\Gamma^4 r\partial_{r}
+\Gamma^5\left(\partial_{\chi} -\frac{1}{2}\Gamma^4\Gamma^5 +i
\right) +i grve^{i\chi}\right\}\Theta(r,\chi)=0.
\end{eqnarray}
Considering $r \ll \sqrt{R_1^2+R_2^2}\ll 1$, we can ignore again
the last item on the RHS of Eq. (\ref{DTCaseII}) and have
$\Theta=f(r)h(\chi)=C \;h(\chi)$, where
$h(\chi)=(h_1(\chi),h_2(\chi))^T$ satisfies the following equation
\begin{eqnarray} \label{hchi}
\left\{ \partial_{\chi} -\frac{i}{2}
\left(%
\begin{array}{cc}
  \gamma^5 & 0 \\
  0 & \gamma^5 \\
\end{array}%
\right) +i \right\}
\left(%
\begin{array}{c}
  h_1(\chi) \\
  h_2(\chi) \\
\end{array}%
\right) =0.
\end{eqnarray}
Imposing the chirality condition $\gamma^5 h_i(\chi) = +h_i
(\chi)$, we get the following solution:
\begin{equation}\label{solution_h}
\Theta(r,\chi)=Ce^{-\frac{i}{2}\chi}
\left(%
\begin{array}{c}
  1 \\
  1 \\
\end{array}%
\right) \otimes \left(
\begin{array}{c}
  1 \\
  0 \\
  1 \\
  0 \\
\end{array}
\right).
\end{equation}
So the fermionic zero mode around the vortex does not depend on
the coupling constant $g$ between extra space and Higgs field and
the vacuum expectation value $v$.

Now we come to the issue of the presence of the zero modes. The
number of zero-modes of the Dirac operator is decided by the index
of it. The index of the Dirac operator on manifold $K^2$ is
defined as the difference $n_{+} - n_{-}$ between the number
$n_{+}$ of right-handed four-dimensional fermions obtained by
dimensional reduction and the number $n_{-}$ of left-handed 4D
fermions. This number is a topological quantity of the manifold
upon compactification and the gauge bundles the Dirac operator
might be coupled to. Indeed, this index can be computed in terms
of characteristic classes of the tangent and gauge bundles. The
Atiyah--Singer index theorem in two dimensions gives the
difference \cite{Atiyah1968_485}
\begin{equation}
n_{+} - n_{-}=\frac{e}{4\pi} \int_{K^2} d^2 q \, \varepsilon^{i j}
F_{i j} \,,
\end{equation}
where $F_{ij}$ is the field strength of $A_{i}$,
\begin{equation}
F_{ij}\equiv \partial_{i}A_{j}-\partial_{j}A_{i} -ie[A_{i}, A_{j}]
\,.
\end{equation}
If we take $K^2=S^2$ with a U(1) magnetic monopole field of charge
$n$ on it, the number of chiral families will then be equal to $n$
\cite{Randjbar1983}. We can also consider zero modes of the Dirac
operator in the background of Abelian gauge potentials
representing Dirac strings and center vortices on the torus $T^2$.
The result is for a two-vortex gauge potential (smeared out
vortices) there is one normalizable zero mode which has exactly
one zero on the torus \cite{Reinhardt2002}. The probability
density of the spinor field is peaked at the positions of the
vortices.

\section{Summary and discussions}

Using the generalized Abelian Higgs model and $\phi$-mapping
theory, we investigate the self-dual vortex on an extra
two-dimensional curved Riemann surface, and obtain the inner
topological structure of the self-dual vortex. We also establish
the exact self-duality equation with topological term, which is
the density of topological charge of the vortex:
$J^{0}=\frac{1}{\sqrt{\gamma}}\frac{2\pi}{e}\delta^{2}(\vec{\phi})J({\phi}/y)$.
Different from the conventional self-duality equation, our
equation with topological term is valid even at the zero points of
the field. In Abelian Higgs model, there are two kinds of vortex:
the topological and non-topological self-dual vortices, which just
arise from the symmetric and asymmetric phase of the Higgs field,
respectively, and are described by a unified equation:
$e\sqrt{\gamma}B=2\pi\sum_{k=1}^NW_k\delta (\vec{y}-\vec{y_{k}})
~\pm ~
\epsilon^{ij}\epsilon_{jk}\partial_{i}(\sqrt{\gamma}\gamma^{kl}\partial_{l}\ln\|\phi\|)
$. As two instances, we give the self-duality vortex equations on
a torus and a sphere.

We give the relation between gauge potential and Higgs field.
Based on this relation, the fermionic zero modes in self-dual
vortex background on a tours are studied. When Higgs filed is only
relative to $\varphi$, the gauge potential $A_{\varphi}$ appears
as phase factor to the fermionic zero modes. So the phase factor
is decided by the gauge potential $A_{\varphi}$. Furthermore, for
the anti-vortex, the corresponding zero mode is singular at the
zero of $\phi$. It suggests that the symmetric phase just
corresponds to singularity of fermionic zero mode, the topological
self-dual vortex just arises from the symmetric phase of Higgs
field. For the vacuum solution, the fermionic zero modes do not
depend on the coupling constant $g$ between extra space and Higgs
field and the vacuum expectation value $v$.

\end{document}